\documentclass[prd,onecolumn,nopacs,nofootinbib,notitlepage,preprintnumbers]{revtex4-1}

\usepackage{mathrsfs}
\usepackage{yfonts}
\usepackage{tipa}
\usepackage{bm,bbm}
\usepackage{bbold}
\usepackage{amsmath}
\usepackage{amssymb}
\usepackage{amsthm}
\usepackage{amsfonts}
\usepackage{mathtools}
\usepackage{color}
\usepackage{latexsym}
\usepackage{relsize}
\usepackage{booktabs}
\usepackage[iso-8859-7]{inputenc}

\usepackage{enumerate}
 \usepackage[colorlinks=true,hyperfootnotes=true,citecolor=cyan]{hyperref}

\usepackage{array}
\newcolumntype{?}{!{\vrule width 2pt}}

\newcommand{\lp}{\left(}
\newcommand{\rp}{\right)}
\newcommand{\lb}{\left[}
\newcommand{\rb}{\right]}

\newcommand{\J}{{{J}}}

\newcommand{\btau}{\boldsymbol{\tau}}

\newcommand{\bV}{\boldsymbol{V}}

\newcommand{\bX}{\boldsymbol{{X}}}

\newcommand{\mL}{{\mathcal L}}

\newcommand{\lsim}   {\mathrel{\mathop{\kern 0pt \rlap
  {\raise.2ex\hbox{$<$}}}
  \lower.9ex\hbox{\kern-.190em $\sim$}}}
\newcommand{\gsim}   {\mathrel{\mathop{\kern 0pt \rlap
  {\raise.2ex\hbox{$>$}}}
  \lower.9ex\hbox{\kern-.190em $\sim$}}}

\newcommand{\ew}{\end{equation}\end{widetext}}
\newcommand{\be}{\begin{equation}}
\newcommand{\ee}{\end{equation}}
\newcommand{\ba}{\begin{eqnarray}}
\newcommand{\ea}{\end{eqnarray}}

\newcommand{\diff}{{{\rm d}}}
\newcommand{\Diff}{{{\rm D}}}

\newcommand{\bt}{\boldsymbol{t}}

\newcommand{\bPhi}{\boldsymbol{{\mathrm{\Phi}}}}
\newcommand{\bPsi}{\boldsymbol{{\mathrm{\Psi}}}}

\newcommand{\bU}{\boldsymbol{{U}}}

\newcommand{\bZ}{\boldsymbol{\mathrm{Z}}}
\newcommand{\bz}{\boldsymbol{\mathrm{z}}}

\newcommand{\bgamma}{\boldsymbol{\gamma}}
\newcommand{\bGamma}{\boldsymbol{\Gamma}}

\newcommand{\bomega}{\boldsymbol{{\omega}}}

\newcommand{\bbe}{\boldsymbol{\rm e}}
\newcommand{\e}{\mathrm{e}}
\newcommand{\ie}{\text{\textschwa}}
\newcommand{\bbie}{\boldsymbol{\textbf{\textschwa}}}

\newcommand{\bkappa}{\boldsymbol{\kappa}}

\newcommand{\bT}{\mathbf{T}}

\newcommand{\bA}{\mathbf{A}}
\newcommand{\bF}{\mathbf{F}}
\newcommand{\bR}{\mathbf{R}}
\newcommand{\bsigma}{\boldsymbol{\sigma}}

\newcommand{\bLambda}{\mathbf{\Lambda}}

\newcommand{\bQ}{\mathbf{Q}}

\newcommand{\nn}{\nonumber}

\newcommand{\Lie}{{\mathcal{L}}}
\newcommand{\nablat}{\hat{\nabla}}
\newcommand{\mE}{{\mathcal{E}}}
\newcommand{\mF}{{\mathcal{F}}}

\begin{document}

\title{The coupling of matter and spacetime geometry}

\author{Jose Beltr{\'a}n Jim{\'e}nez}
\affiliation{Departamento de F{\'i}sica Fundamental and IUFFyM, Universidad de Salamanca, E-37008 Salamanca, Spain}
\email{ jose.beltran@usal.es}
\author{Lavinia Heisenberg}
\affiliation{Institute for Theoretical Physics, ETH Zurich, Wolfgang-Pauli-Strasse 27, 8093, Zurich, Switzerland}
\email{lavinia.heisenberg@phys.ethz.ch}
\author{Tomi Koivisto}
\affiliation{Laboratory of Theoretical Physics, Institute of Physics, University of Tartu, W. Ostwaldi 1, 50411 Tartu, Estonia}
\affiliation{National Institute of Chemical Physics and Biophysics, R\"avala pst. 10, 10143 Tallinn, Estonia}
\affiliation{Helsinki Institute of Physics, P.O. Box 64, FIN-00014 Helsinki, Finland}
\affiliation{Department of Physical Sciences, Helsinki University, P.O. Box 64, FIN-00014 Helsinki, Finland}
\email{tomik@astro.uio.no}

\date{\today}

\begin{abstract}

The geometrical formulation of gravity is not unique and can be set up in a variety of spacetimes. Even though the gravitational sector enjoys this freedom of different geometrical interpretations, consistent matter couplings have to be assured for a steady foundation of gravity. In generalised geometries, further ambiguities arise in the matter couplings unless the minimal coupling principle (MCP) is adopted that is compatible with the principles of relativity, universality and inertia.  In this work, MCP is applied to all Standard Model gauge fields and matter fields in a completely general (linear) affine geometry. This is also discussed from an effective field theory perspective. It is found that the presence of torsion generically leads to theoretical problems. However, symmetric teleparallelism, wherein the affine geometry is integrable and torsion-free, is consistent with MCP. The generalised Bianchi identity is derived and shown to determine the dynamics of the connection in a unified fashion.  Also, the parallel transport with respect to a teleparallel connection is shown to be free of second clock effects. 
\end{abstract}

\maketitle

\tableofcontents

\section{Introduction}

Reference frames in Special Relativity are symmetric under the global ISO(1,3) transformations and standard particles live in representations of that group. The gravitational interaction of General Relativity is switched on by localising the symmetry, as was realised in Kibble's construction of Poincar\'e gauge theory of gravity \cite{Kibble:1961ba}. In this process, the actions $I$ of the standard model fields $\phi$ are made locally invariant by promoting the metric of the inertial frame to the dynamical spacetime metric $\eta \rightarrow g$ and replacing the partial derivatives with the covariant ones $\partial \rightarrow \nabla$. This is  
\begin{itemize}
\item[] {\it the minimal coupling principle } : \quad $I(\eta,\phi,\partial\phi) \rightarrow   I(g,\phi,\nabla\phi)$ \quad\quad\quad\quad\quad\quad\quad\quad\quad (MCP) 
\end{itemize}
concerning any relevant collection of fields $\phi$ but restricted to the unique inner product $\eta$ given by the symmetry group G of the connection $\nabla$, and which should be distinguished from 
\begin{itemize}
\item[] {\it the metrical coupling principle  } : \quad $I(\eta,\phi,\partial\phi) \rightarrow   I(g,\phi,\mathcal{D}\phi)$ \quad\quad\quad\quad\quad\quad\quad\quad\quad (mCP) 
\end{itemize}
where $\mathcal{D}=\mathcal{D}(g,\partial g)$ is the unique torsion-free connection compatible with $g$, and which is just one of the infinite number of possible well-defined but non-minimal prescriptions unless $\mathcal{D} = \nabla$. 

The issue of minimal coupling is of paramount importance in the current discussions of alternative (gauge) theories of gravity. In particular, the G=ISO(1,3)=SO(1,3)$\rtimes\mathbbm{R}^4$ symmetry can be extended to the inhomogeneous general linear symmetry G=IGL=GL(4,$\mathbbm{R}$)$\rtimes\mathbbm{R}^4$ if $\phi$ includes infinite-dimensional matrices \cite{Hehl:1994ue} or simply the homogeneous general linear symmetry G=GL(4,$\mathbbm{C}$) that accommodates standard spinors \cite{Koivisto:2019ejt}. In such contexts, one often classifies affine connections according to whether they are flat, torsion-free and metric-compatible. In 7 of the resulting 8 distinct classes of geometries, it is possible to construct gravitational actions that reproduce the dynamics of General Relativity, giving rise to for example the Geometrical Trinity \cite{BeltranJimenez:2019tjy} (see also \cite{Heisenberg:2018vsk}), and the recently introduced generalisation of teleparallel geometries \cite{Jimenez:2019ghw}.  -- Alternative formulations for the equivalent classical dynamics abound, the question arises whether the geometry of spacetime can be decided by experiments, or whether it is merely a matter of convention. Whilst the latter may be the case for the gravity action in vacuum\footnote{This might be compared with the well-known case of conformal frames in scalar-tensor theory \cite{Brans:1961sx}, wherein the dynamics can be equivalently described in terms of an arbitrarily rescaled metric, but only in the so called Jordan frame that is distinguished by the minimal matter coupling, do those dynamics have their standard physical interpretation (i.e. assuming that test particles follow geodesics, constants of nature do not vary, etc.) \cite{Koivisto:2005yk}.}, contrasting it with matter may give a unique answer if MCP is adopted \cite{So:2006pm}. This is the motivation of this paper. 
 
In metric teleparallel gravity, the coupling of spin-$\frac{1}{2}$ fields was some time ago an issue of some controversy in the literature \cite{Obukhov:2002tm,Maluf:2003fs,Mielke:2004gg,Obukhov:2004hv}. While it is generally agreed that the gravitational coupling of spinors to the metric teleparallel connection (Weitzenb\"ock $mod$ pure gauge) is inconsistent, the problem is avoided by invoking MCP.  
This is sometimes advocated as the {\it teleparallel coupling prescription}\footnote{One may always define the distortion $X \equiv\mathcal{D}-\nabla$ (as will be clarified by (\ref{decomposition})). In effect, the coupling prescription amounts to the trivial rewriting of mCP as mCP': $I(\eta,\phi,\partial\phi) \rightarrow   I(g,\phi,(\nabla + X)\phi)$. The (metric) teleparallel theory may offer a specific rationale for this non-minimal coupling prescription \cite{deAndrade:1997cj,Aldrovandi:2013wha,Krssak:2018ywd}, but it can obviously be applied for any $X$.} mCP', which has been stipulated for electromagnetic fields \cite{deAndrade:1997cj}, scalars \cite{deAndrade:1997gka,deAndrade:2001vx} and spinors \cite{Maluf:2003fs,Mosna:2003rx}. MCP' has been arrived at also in symmetric teleparallelism \cite{Adak:2011rr} and was recently exploited with a generic $\nabla$ \cite{Delhom:2020hkb}. Though the coupling mCP' is a mathematically well-defined prescription, we reiterate that there is no ambiguity of the gauge principle MCP in the standard framework of Yang-Mills theory, where the $\phi$ consists of sections to associated bundles, transformations of which are canonically determined in conjunction with the transformation of the $\nabla$ on the principal G-bundle. One may thus share the sentiment of the footnote 2 of \cite{Itin:2018dru}. An important remark in this respect is that one is left with a choice to make concerning the undetermined charge of the matter fields\footnote{In multi-field theories it might be necessary to require some non-trivial relations among the different charges even before introducing the couplings to the gauge fields. For instance, a theory with three scalar fields $\phi_1$, $\phi_2$, $\phi_3$ with an invariance under $\phi_n\rightarrow e^{iq_n\alpha}\phi_n$, with $q_n$ the corresponding charges and $\alpha$ the transformation parameter, permits an interaction such as $\phi_1\phi_2\phi_3$ provided the charges satisfy the relation $q_1+q_2+q_3=0$. However, only the coupling to the $U(1)$ gauge field will eventually determine the precise values of the individual charges $q_n$.}. In other words, matter fields sharing the same dynamics in the absence of gauge fields can be differentiated by their interactions with them.  As we will discuss in more detail below, this issue also applies to gravity. However, if we embrace the equivalence principle and stipulate the universal character of gravity, it is possible to avoid this ambiguity and establish a unique MCP for gravity.

In symmetric teleparallel gravity \cite{Nester:1998mp}, MCP is viable for all standard fields, including spinors \cite{BeltranJimenez:2017tkd}. The Hermitean Dirac action, minimally coupled to a symmetric teleparallel (coincident $mod$ pure gauge \cite{BeltranJimenez:2017tkd,Koivisto:2018aip}) connection lets spinors interact only with the metrical connection, and in the case of complex parameters, a phase gauge field \cite{Koivisto:2018aip}. In this note, we confirm and generalise these results. 

It is crucial to note that MCP concerns the actions $I$ (and not the field equations $\delta I/\delta\phi$) in order to establish the consistency of the Coincident General Relativity \cite{BeltranJimenez:2017tkd, Koivisto:2018aip}. In general, when the two prescriptions are inequivalent, it is because the alternative to MCP could only be derived from a non-Hermitean $I$, and resulted therefore in either a non-unitary or non-conservative theory. It is also intuitively clear that the alternative does not lead to physical results, since it would e.g. couple the massless Dirac theory to a scale connection, even though the theory is conformally invariant. Since we are not aware of a proper justification for the alternative, we shall not discuss it further\footnote{Nevertheless, it could be possible to meaningfully constrain the non-Hermitean coupling \cite{Soo:1997gp}.}.

In this paper we shall work out the implications of MCP with an arbitrary affine connection.
We will show in Section \ref{gaugefields} that MCP is already problematic for bosonic fields\footnote{At the risk of resulting pedantic, it may be worthwhile to clarify the terms bosonic and fermionic in a general framework. These terms are traditionally borrowed from the transformation properties of fields under the Lorentz group depending on whether they belong to some tensor product of vector representations (bosons) or they live in the universal (double) cover of SO(1,3). Extending these properties to a more general G group can be subtle and, as a matter of fact, this is a source of complications for dealing with fermions in a general scenario as we will clarify below.} if the connection has non-vanishing torsion.  In Section \ref{inner} we clarify the geometrical meaning of non-metricity. The action $I(g,\phi,\nabla\phi)$ for fermions, which was derived \cite{Koivisto:2018aip} using the Hermitean map on the GL(2,$\mathbbm{C}$)$\otimes$GL(2,$\mathbbm{C}$) bundle, is rederived in Section \ref{spinor1} on the GL(4,$\mathbbm{C}$) bundle and yet presented in Appendix \ref{reps} on its double-covering 
SU(2,2) bundle: all three realisations can yield the same coupling of spinors to spacetime geometry. 
In Section \ref{spinor2} we use a generalised Noether identity from Appendix \ref{noether}, derive the equations of motions for $g$, $\nabla$ and $\phi$,  and clarify their implications in the main cases of interest. Especially determinant to these implications is the hypermomentum structure of the connection equations $\delta I/\delta\nabla$. We conclude in Section \ref{concs}.

\section{On MCP from an EFT perspective}
\label{eft}

Before delving into the core of the main topic of our study, we will briefly discuss the role of MCP from a pure effective field theory (EFT) perspective as well as its physical necessity from this viewpoint. MCP is usually advocated as the appropriate prescription to couple gauge fields to a matter sector that is charged under the corresponding group. It will then be instructive to commence our discussion by briefly reminding how this prescription comes about. 

For the sake of simplicity, let us focus on the case of an Abelian U(1) gauge field $A_\mu$ that transforms as $\bA\rightarrow  \bA+\diff\theta$ with $\theta$ some arbitrary 0-form. The most important physical property of this field is its masslessness which in turn represents the underlying reason for introducing the gauge redundancy that guarantees the propagation of two polarisations. The standard procedure to guarantee this gauge symmetry is by constructing its action in terms of its field strength $\bF=\diff\bA$. If the gauge field is to couple to some matter sector, the interactions must respect the gauge symmetry to avoid the appearance of undesired new polarisations. The important question arises as how to introduce couplings to matter.

If the matter sector does not feature any conserved current, we are limited to derivative couplings where the gauge field only enters through its field strength and the symmetry is trivially realised by only transforming the gauge field. Examples of this type of couplings are provided by e.g. a dilaton $\varphi$ or an axion $\vartheta$ that couple to the electromagnetic field like $\varphi F_{\mu\nu} F^{\mu\nu}$ and $\vartheta F_{\mu\nu} \tilde{F}^{\mu\nu}$ respectively. For a fermion $\psi$ we similarly have the Pauli interaction $\bar{\psi}[\gamma^\mu,\gamma^\nu]\psi F_{\mu\nu}$ that respects all the desired symmetries of the theory. In all these cases, the interactions could have not been guessed nor constructed by resorting to the U(1) minimal coupling prescription, but obviously there is nothing wrong with them and, in fact, they all are present in the corresponding theories. Let us then see the relevance of the minimal coupling prescription within an EFT approach.

If the matter sector does carry a conserved current, there is another class of interactions that do not involve derivatives of the gauge field and where the realisation of the gauge symmetries involves transforming the matter sector along with the gauge field. These interactions can be constructed iteratively as an expansion on a coupling constant and whose resummation precisely  corresponds to the application of the minimal coupling prescription (see e.g. the nice discussion in \cite{ortin_2004}). Of course, this is nothing but the localisation of the global symmetry leading to the conserved current. To give an explicit, nearly trivial example, we can resort to the well-known case of scalar electrodynamics where, starting from a complex scalar field $\Phi$ with a global symmetry $\Phi\rightarrow e^{iq\alpha}\Phi$, it can be coupled via MCP that localises this global symmetry as part of the U(1) symmetry of the gauge field. This procedure leads to interactions such as $\Phi^*A^\mu\partial_\mu\Phi$ and $\vert\Phi\vert^2A^2$ with very specific coefficients dictated by gauge symmetry. The fact that these interactions are lower order in derivatives than those involving the field strength makes it clear that the former will conform the relevant interactions at low energies. A direct consequence of this is the $1/r$ asymptotic fall-off generated by the corresponding sources that gives the long-range interaction expected for a massless field. Nothing prevents from adding interactions such as $\vert\Phi\vert^2F^2$ which are not generated via MCP, but these will be suppressed by some energy scale that will make it perturbatively unimportant as compared to e.g. $\vert\Phi\vert^2A^2$. For instance, the field generated by a source would decay faster than $1/r$. It is remarkable that all these properties (gauge symmetry, conservation laws, minimal coupling...) root in the very masslessness of fields and can be nicely derived from processes involving external soft massless particles (\cite{Weinberg:1995mt}). Concerning massless spin-2 fields, the universal character of gravity encoded in the equivalence principle can then be derived as a necessary property of the leading order interactions.

Thus, it could be argued that MCP is the appropriate prescription to generate the most relevant interactions at the lowest order in an energy expansion, i.e., within an EFT framework, and this would sustain referring to this prescription as a principle. However, one should keep in mind the underlying reason for invoking this coupling prescription and decide if one wants to uplift it to the category of {\it fundamental principle}. Similarly, the universality of the coupling to gravity is an accidental property of the leading order interactions from an EFT perspective. Hence, imposing universality for the couplings to gravity signifies to promote it to the category of a more fundamental principle that steps outside the realm of EFTs.

As we will see in Section \ref{gaugefields}, one important consequence of adopting the minimal coupling prescription $\diff\rightarrow \nabla$ is that a direct coupling of the torsion to the gauge field is generated so that the very defining property of the gauge field as a massless field is lost\footnote{Therefore one may reconsider beginning the EFT construction in a contorted space, which would result in the elimination of the direct coupling realising mCP'.}.
After all, the gauge invariance is invoked precisely to maintain the masslessness of the fields. It may be convenient to recall here that it is precisely the consistency of the lowest order interactions between the massless spin-1 fields what forces upon us the underlying gauge structure that dictates how the different fields must interact. It is this requirement what associates massless fields to connections in a principal bundle and, furthermore, tells us that the interactions between connections must occur in compliance with some gauge structure. Thus, it would seem natural to conclude that a direct, non-derivative coupling between a gauge field and the torsion could only happen if they belonged to non-commutative sectors of some gauge group.

Let us be a bit more explicit on this point by taking a specific example with a set of $N$ Abelian gauge fields $A^a{}_\mu$ with their corresponding field strengths $F^a=\diff A^a$. We will have in mind that some of these fields will eventually become the general linear connection, but for the moment we shall not require anything. The free propagation of these fields will be described by the usual Lagrangian
\be
\mL_{\rm free}=-\frac14 \mathcal{M}_{ab} F^a{}_{\mu\nu} F^{b\mu\nu}.
\ee
with $\mathcal{M}_{ab}$ some metric in the internal space of the fields. If this metric has some isometries, these will give rise to a set of on-shell conserved currents. Reversely, this internal metric can be chosen as the Killing metric of the internal symmetry group that we may demand. The next step is to introduce interactions among the different gauge fields. At the lowest order, we would seek to introduce non-derivative couplings, but, as it is well-known, keeping the masslessness of all the gauge fields comes hand-in-hand with strict restrictions on the allowed interactions. For instance, one cannot have a cubic interaction with only two gauge fields, but at least three are necessary. A systematic construction of the possible interactions can be carried out as a perturbative series in the coupling constant and, at the end of the process, one finds that the global symmetry and the independent $U(1)$ gauge symmetries combine to give a gauge symmetry associated to some non-Abelian group and the interactions are precisely dictated by the non-Abelian structure.  
The problems root in the gauge fields belonging to the adjoint representations so they transform as connections and this is at the heart of the very nature of massless fields with spin higher than zero.

In the precedent paragraphs we have not really entered into the quantum domain and remained at the classical level. It is interesting to emphasise as well that MCP will be {\it violated} via quantum effects, i.e., even if we stick to the interactions prescribed by MCP at the classical level, new interactions not complying with MCP will be generated quantum mechanically and this can be originated from two sources. 

On one hand,  quantum corrections are expected to generate interactions beyond MCP, because, even if the tree amplitudes conform to MCP, loops can generate interactions that violate MCP. It is crucial however that the theory is interpreted as an EFT so that non-renormalisable operators are present. If the theory is renormalisable, then it is obvious that MCP is stable under the quantum corrections. A paradigmatic example of MCP violation within the context of gravity is given by the EFT interpretation of General Relativity. If we apply a MCP procedure to a canonical scalar field, then, at the one-loop level, direct couplings of the scalar field to the curvature will be generated. Ultimately, this is because loop processes permit the scalar field to sniff around the spacetime so it will be sensitive to its geometrical properties. On the other hand, it is well-known that classical symmetries can be broken at the quantum level via anomalies when the path integral measure or the renormalisation scheme do not respect them. Paradigmatic examples of this kind are the breaking of scale invariance or the axial anomaly that, for instance, crucially permits the decay of the neutral pion into two photons via the celebrated triangle diagram. This experimentally observed process would be forbidden had the minimal coupling prescription taken the status of a fundamental principle.

Our above discussion puts forward that MCP seems to be the appropriate prescription in order to generate the leading order interactions in a given theory, but from an EFT perspective there is no fundamental reason why only interactions complying with MCP should be considered.


\section{Gauge fields}
\label{gaugefields}

Though it takes just a one-line-calculation to arrive at our conclusion using just the electromagnetic field, in this Section we set up the notation by generalising the derivation to arbitrary gauge fields (external and internal, assuming direct product) and by studying also their Bianchi identities.

\underline{{\bf Lemma 1.}}  Consider the gauge field ${\bLambda}$ and the gauge-covariant exterior derivative $\Diff$. {\bf a}) We can write the canonical field strengths ${\bf F}$ as ${\bf F}=\Diff {\bLambda}$, iff the connection has no torsion. Furthermore,  {\bf b)} the Bianchi identity can then and only then be written as $\Diff {\bf F}=0$. 
\newline
\newline
{\bf Proof:}  We expand the gauge field ${\bLambda}=\Lambda^N \J_N$ in the basis of generators $\J_N$ that satisfy the commutation relations $[\J_K,\J_L]=f^N{}_{KL}\J_N$, with $f^N{}_{KL}$ the structure constants. 
Let the general linear part of $\Diff$ be denoted by $\nabla=\diff + \Gamma$, so that $\Diff=\nabla + [\bLambda,\,\,\,]$. Our conventions are such that for vector components $V^\mu$ and for one-form components $W_\mu$,
\begin{subequations}
\label{covariants}
\ba 
\nabla_\mu V^\alpha & = & V^\alpha{}_{,\mu} + \Gamma^\alpha_{\phantom{\alpha}\mu\lambda}V^\lambda\,, \\
\nabla_\mu W_\alpha & = & W_{{\alpha},\mu} - \Gamma^\lambda_{\phantom{\alpha}\mu\alpha}W_\lambda\,,
\ea
\end{subequations}
respectively. It then follows that 
\begin{subequations}
\label{geometry}
\ba 
\big[\nabla_\mu,\nabla_\nu\big]V^\alpha & = & {R}^\alpha{}_{\beta\mu\nu}V^\beta - {T}^\beta{}_{\mu\nu}\nabla_\beta V^\alpha\,, \\
\big[\nabla_\mu,\nabla_\nu\big]W_\alpha & = & -{R}^\beta{}_{\alpha\mu\nu}W_\beta - {T}^\beta{}_{\mu\nu}\nabla_\beta W_\alpha\,, \label{geometry2}
\ea
\end{subequations}
which define 
\begin{subequations}
\ba
\text{the curvature}: \quad  {R}^\alpha{}_{\beta\mu\nu}  & = &  2\partial_{[\mu} \Gamma^\alpha{}_{\nu]\beta} 
+ 2\Gamma^\alpha{}_{[\mu\lvert\lambda\rvert}\Gamma^\lambda{}_{\nu]\beta}\,, \\ \text{and the torsion:} \quad  {T}^\alpha{}_{\mu\nu}  & = &  2\Gamma^\alpha{}_{[\mu\nu]}\,, 
\ea
\end{subequations}
of the connection $\Gamma$. Their derivatives satisfy the purely geometric identities 
\begin{subequations}
\ba
R^\mu_{\phantom{\mu}[\alpha\beta\gamma]} - \nabla_{[\alpha}T^\mu_{\phantom{\mu}\beta\gamma]} + T^\nu_{\phantom{\nu}[\alpha\beta}T^\mu_{\phantom{\mu}\gamma]\nu} & = & 0\,, \label{Bianchi2}\\
\nabla_{[\alpha}R^{\mu}_{\phantom{\mu}\lvert\nu\rvert\beta\gamma]} - T^\lambda_{\phantom{\lambda}[\alpha\beta}R^{\mu}_{\phantom{\mu}\vert\nu\rvert\gamma]\lambda} & = & 0\,,  \label{Bianchi3}
\ea
\end{subequations}
that can be directly obtained from the Jacobi identity applied to $\nabla$ acting on a vector field. We can now be very explicit and write 
\ba
\Diff {\bLambda} = \Diff_{[\mu}\Lambda^N{}_{\nu]}\J_N\diff x^\mu\wedge \diff x^\nu   & = &   \lp \nabla_{[\mu}\Lambda^N{}_{\nu]} + f^N{}_{KL}\Lambda^K{}_\mu \Lambda^L{}_\nu \rp \J_N\diff x^\mu\wedge \diff x^\nu \nonumber \\ & = & 
 \lp \partial_{[\mu}\Lambda^N{}_{\nu]} - T^\alpha{}_{\mu\nu}\Lambda^N{}_\alpha +  f^N{}_{KL}\Lambda^K{}_\mu \Lambda^L{}_\nu \rp \J_N\diff x^\mu\wedge \diff x^\nu 
 \nonumber \\ & = & {\bf F} - T^\alpha{}_{\mu\nu}{\bLambda}_\alpha\diff x^\mu\wedge \diff x^\nu\,,
\ea 
which proves {\bf 1a)}. For the second part, let us use exterior algebra, wedging together the bold symbols 
\be
\Diff \bF  =  \nabla \bF + [\bLambda,\bF] = \nabla\lp \nabla\bLambda+\bLambda^2\rp + [\bLambda, \nabla\bLambda+\bLambda^2] = \nabla^2 \bLambda + [\nabla\bLambda,\bLambda] + [\bLambda,\nabla\bLambda] = \nabla^2\bLambda\,. 
\ee
At this point we have assumed the Jacobi identity $f^D{}_{[AB}f^E{}_{C]D}=0$ of a Lie algebra so that $\bLambda^3=0$.
Recalling (\ref{geometry2}) and then using (\ref{Bianchi2}), we obtain that
\ba
\Diff \bF & = &  -\lp {R}^\beta{}_{\alpha\mu\nu}\bLambda_\beta + {T}^\beta{}_{\mu\nu}\nabla_\beta \bLambda_\alpha\rp \diff x^\alpha\wedge\diff x^\mu\wedge\diff x^\nu
\nonumber \\
& = &  \lp T^\lambda{}_{\alpha\mu}T^\beta{}_{\gamma\lambda}\bLambda_\beta - \nabla_\alpha T^\beta{}_{\mu\nu}\bLambda_\beta - T^\beta{}_{\mu\nu}\nabla_\beta \bLambda_\alpha\rp
\diff x^\alpha\wedge\diff x^\mu\wedge\diff x^\nu\,. \label{result}
\ea 
 {\bf 1b)} is proven. We note that the necessary and sufficient condition for the $\nabla$ to obey the Poincar\'e lemma is that the $\Gamma$ is symmetric. 

The above Lemma shows the problem of metric teleparallelism already for bosonic fields if one adopts MCP. 
Even standard fields as the photon will be forced to couple to the torsion non-minimally, which further jeopardizes the standard form of the U(1) invariance\footnote{A modified U(1) transformation has been proposed to accommodate one scalar degree of freedom in torsion \cite{Hojman:1978yz}.}. 
Since the algebraic structure 
\be
\Diff\bLambda = \nabla\bLambda + \frac12[\bLambda,\bLambda] =  \nabla\bLambda + \bLambda\wedge\bLambda = \bF - (\bbie_a\cdot\bLambda)\bT^a\,,
\ee
is reflected in the spacetime geometry such that precisely torsion in affine geometry fails to preserve it, 
in symmetric teleparallelism the problem is avoided elegantly since torsion is zero and one has ${\bf F}=\Diff {\bLambda}$ as in pure Riemann geometries.

Let us mention that there is a simple way to avoid the problem, though by violating MCP and giving up strictly the universality of gravity. One could consider the coupling of total gauge connection, consisting of the direct sum of the GL and the internal connection.      
In other words, we could take $\Gamma+\bLambda$ as the full connection so that the field strength would be
\be
\bF_{\rm total}=[\Diff_{\rm total},\Diff_{\rm total}]=\diff\big(\Gamma+\bLambda)+[\Gamma+\bLambda,\Gamma+\bLambda]=\bR+\bF\,, 
\ee
where we have used that $[\Gamma,\bLambda]=0$, since they belong to different subspaces. Since $\bLambda$ is not a section on a bundle associated with the spacetime connection $\Gamma$, but together with $\Gamma$ a part of the connection on the principal bundle, we could call this the Unified Coupling Principle (UCP), defined by the different rules in the gauge and in the matter sectors. Since (UCP) by construction preserves the given symmetries in both sectors, it is also compatible with the EFT perspective of Section \ref{eft}.

Although we have obtained these results for non-Abelian gauge fields described by 1-forms, it is straightforward to extend it to arbitrary $p$-forms. In general, for a given $p$-form $\bA_p$ it holds the identity
\be
\nabla_{[\mu}A_{\nu_1\cdots\nu_p]}=\partial_{[\mu}A_{\nu_1\cdots\nu_p]}-p\,T^\alpha{}_{[\mu\nu_q}A_{\nu_1\cdots\nu_{q-1}]\alpha}\,,
\ee
so our discussions are applicable to general $p-$form fields as well. In four dimensions, massless 2-forms are dual to 0-forms and massless 3-forms are non-dynamical so our discussion on 1-forms is exhaustive. In higher dimensions however there is a richer landscape and analogous shortcomings should be  considered.


\section{Parallel transported clocks}
\label{inner}

This far we have discussed only a connection (in both internal and spacetime geometry) which sufficed for (MPC) with gauge fields. To discuss the half-integer spin matter fields, it is necessary to introduce also a metric. But before moving to matter fields, let us consider a related issue that arises in the presence of both metric and affine structure. In particular, they may be incompatible, in which case the usual metrical concepts may not be uniquely defined for parallel transported objects. 

However, we emphasise that the problem is of no direct relevance to the behaviour of matter fields. As suggested by the extremisation of the proper time of a point particle, matter tends to follow the metric geodesics regardless of an independent connection; our study of spinor fields will confirm (at least in symmetric teleparallelism) this suggestion from first principles. Thus, the physical relevance of the evolution of a metric contraction during parallel transport with respect to a non-metric connection is not so immediately clear. It is, nevertheless, a very basic aspect of metric-affine geometry and thus worth clarification.

Take, as usual, a metric tensor with the components $g_{\mu\nu}$.
Then the incompatibility of the affine connection is characterised by the non-metricity tensor $Q_{\alpha\mu\nu} = \nabla_\alpha g_{\mu\nu}$. We shall prove the following 
\newline
\newline
\underline{{\bf Lemma 2.}} The inner product is path-independent iff $\bR_{(ab)}=0$. 
\newline
\newline
 {\bf Proof}: Consider two vector fields $\bU,\bV$  parallel transported along a curve $\gamma$ with the tangent vector $\bX$. The change of the inner product $(\bU,\bV)=g_{\mu\nu}U^\mu V^\nu$ along the curve is obviously given by $\nabla_{\bX} (\bU,\bV) = Q_{\mu\alpha\beta} X^\mu U^\alpha V^\beta$. We take $\gamma$ to be a closed curve, a loop, since it is relevant to issues such as the second clock effect which require observers to compare notes. The total change is given by integrating $\nabla_{\bX} (\bU,\bV)$ around the $\gamma$, and by the Stokes'  theorem {\cite{Eguchi:1980jx}}
it becomes an integral over a surface $S$ outlined by $\gamma$,
\be
\Delta (\bU,\bV) = \oint_\gamma Q_{\mu\alpha\beta} U^\alpha V^\beta\diff x^\mu = \iint_S \partial_{[\mu}\lp  Q_{\nu]\alpha\beta} U^\alpha V^\beta\rp \diff x^\mu\wedge \diff x^\nu\,.  
\ee 
By substituting the covariant derivative we obtain
\be \label{integral}
\Delta (\bU,\bV) = \iint_S\lb \lp \nabla_{[\mu}Q_{\nu]\alpha\beta} + \frac{1}{2}T^\lambda{}_{\mu\nu}Q_{\lambda\alpha\beta}\rp U^\alpha V^\beta 
+ \nabla_{[\mu}\lp U^\alpha V^\beta\rp Q_{\nu]\alpha\beta}\rb \diff x^\mu\wedge \diff x^\nu\,. 
\ee
Using the third (metric) Bianchi identity \cite{schouten1954ricci,BeltranJimenez:2018vdo}
\be
\nabla_{[\mu}Q_{\nu]\alpha\beta} = -R_{(\alpha\beta)\mu\nu} - \frac{1}{2}T^\lambda{}_{\mu\nu}Q_{\lambda\alpha\beta}\,,
\ee
we see that the term $\sim U^\alpha V^\beta$ in (\ref{integral}) is proportional to the symmetric part of the curvature. The remaining term $\sim \nabla (U^\alpha V^\beta)$ in (\ref{integral}) can be set to zero for the parallel transported vector fields upon the chosen surface. We have thus arrived at
\be \label{integral2}
\Delta (\bU,\bV) = -\iint_S R_{(\alpha\beta)\mu\nu} U^\alpha V^\beta\diff x^\mu\wedge \diff x^\nu\ = -2\iint_S \bR_{(ab)} U^a V^b\,.
\ee
Lemma {\bf 2} is verified. This result can be derived in a more straightforward manner by resorting to exterior calculus:
\be
\Delta (\bU,\bV) = \oint_\gamma Q_{ab} U^a V^b =\iint_S \diff\Big[Q_{ab} U^a V^b\Big]=\iint_S \Diff\Big[Q_{ab} U^a V^b\Big]=\iint_S \Diff Q_{ab} U^a V^b=-2\iint_S \bR_{(ab)} U^a V^b\,,
\ee 
where we have used the Bianchi identity $\Diff Q_{ab}=-2\bR_{(ab)}$ and the parallel transport condition.

One immediate implication of this result is that in parallel transported objects in teleparallel spacetimes (symmetric or otherwise) do not experience a second clock effect (as was already stated without proof in \cite{BeltranJimenez:2019tjy}, but contrary statements are also found \cite{Delhom:2020vpe}). Indeed, the geometrical foundation of ``purified gravity'' is a generalisation \cite{Koivisto:2018aip} of a Weyl integrable spacetime (WIST) \cite{Brans:1961sx,Rosen:1982nr,Scholz:2019tif}. 
In a general Weyl spacetime, $Q_{\alpha\mu\nu}=\frac{1}{4}Q_\alpha g_{\mu\nu}$,  and thus $\nabla_{\bX} (\bU,\bV) = \frac{1}{4}(\bU,\bV) \bQ(\bX)$, yielding immediately the well-known result $\Delta \log{(\bU,\bV)} =  -\frac{1}{4}\iint_S \bR^a{}_a$.  The vanishing of the Streckenkr\"ummung a.k.a. homothetic curvature\footnote{Actually, this component corresponds to the 
overall, direction-independent change of scale, while the rest of the symmetric curvature describes how shapes are "sheared" or "disformed" through both local rotation and variation in lengths.} $\bR^a{}_a=0$ in a WIST wherein $\bQ=\diff Q$ for some scalar $Q$,  guarantees the path-independence of the inner product. 


\section{Matter fields}

In Section \ref{gaugefields} we have discussed the case of gauge fields separately because of their special status and properties which are tightly related to their masslessness. We turn our attention now to the {\it matter sector}. Our distinction closely follows the usual classification of particle physics where gauge fields are associated to interactions. In the matter sector we can distinguish two crucially different classes of matter fields: bosons and fermions. As we will discuss, bosons can be easily coupled to gravity, but fermions are more subtle. 

The description of fermions in the presence of gravity is substantially more contrived and subtle than for bosonic fields. The underlying reason for the additional complications resides in the fact that bosonic fields are described by tensor representations while fermions require spinor fields. The starting point to introduce gravity is the flat spacetime version of the theory endowed with a Lorentzian structure. When switching on gravity, Lorentz tensors become GL$(4,\mathbbm{R})$ -tensors univocally through the soldering form so no ambiguity arises and one can straightforwardly map the SO(1,3)-connection in the Lorentz bundle to an affine connection in the GL$(4,\mathbbm{R})$ -bundle. For spinors however this is not a direct procedure because it first requires obtaining the universal (double) cover of the Lorentz group and the direct translation to GL$(4,\mathbbm{R})$  is, in general, not possible. In other words, unlike for tensor representations, there is no isomorphism for the corresponding spinor representations. In fact, constructing spinor representations for GL$(4,\mathbbm{R})$  is by itself a non-trivial task. This lack of an isomorphic relation between spin representations introduces an obstruction for the definition of the corresponding connection. It is possible to trace the main difficulty to the presence of non-metricity that obstructs to map the spin connection associated to the Lorentz bundle (more precisely, the connection in Spin(1,3) $\simeq$ SL$(2,\mathbbm{C}$)) to the GL$(4,\mathbbm{R})$  bundle. In the absence of non-metricity, it is possible to use the Kosmann lift to establish the desired map. For this reason, we will carefully derive our results for fermions below, but let us first briefly consider the simplest bosonic fields.

\subsection{Bosonic fields}
Bosonic fields are described by Lorentz tensors in the starting inertial theory without gravity. As we said above, the isomorphic correspondence between tensor representations of SO(1,3) and GL$(4,\mathbbm{R})$ eases the introduction of their couplings to gravity with the covariant derivative complying with the MCP. It is worth however to mention some subtle points that might arise. Firstly, although there is an isomorphism for the tensor representations, there is no way of distinguishing between (proper) tensor densities of different weights for the Lorentz group. The (pseudo-)orthogonal nature of the Lorentz transformations trivialises the weight dependence of tensor representations, the only important property being their behaviour under parity.  When turning on gravity, the weight of the tensor densities matters and the covariant derivative sees it, i.e., it includes an additional contribution to correct for the weight. Thus, we need to make a choice for the weight when promoting the Lorentz tensors to their curved versions.

It is also interesting to emphasise what happens for massless gauge fields that further motivates the separate dedicated discussion in Section \ref{gaugefields}. In order to be specific, let us consider again a massless spin-1 field. It is then well-known that its polarisation vector does not transform as a Lorentz vector under Lorentz transformations, but it picks an inhomogeneous part. A consequence of this anomalous transformation is that the operator describing the gauge field $A_\mu$ transforms under a Lorentz rotation parameterised by $L^\alpha{}_\beta\in$ SO(1,3) as $A_\mu\rightarrow \Lambda_\mu{}^\nu A_\nu+\partial_\mu\Omega$, with $\Omega$ an arbitrary function. This does not correspond to how a Lorentz vector transforms so that mapping it to a GL$(4,\mathbbm{R})$-vector is not possible. The difficulty can be easily solved by assuming that the homogeneous part is mapped to the GL$(4,\mathbbm{R})$-version while the inhomogeneous part remains the same. This observation shows another view on the specific troubles for gauge fields that complements those already explained in \ref{gaugefields}. In particular, since it does not transform as a tensorial quantity, defining a covariant derivative can be ambiguous. Of course, the physical quantity is given by the corresponding field strength for which the inhomogeneous part drops and, therefore, it does transform as a tensor.

After briefly commenting on the potentially ambiguous points of applying the MCP to bosonic fields, let us delve into the more subtle case of fermions.

\subsection{Fermionic fields}
\label{spinor1}

We will start by stating the following
\newline

\underline{\bf Lemma 3.} Consider MCP in the Hermitean theory of Dirac. The action is unaffected by a real, affine generalisation of the metric connection iff the generalised connection has no axial torsion.   
\newline
\newline
{\bf Proof:} In the more subtle case of fermions it is pertinent to report the derivations in greater detail. Though irrelevant for the Lemma {\bf 2},  for generality we consider the connection of the complexified General Linear group GL(4,$\mathbbm{C}$)\footnote{This may be convenient because of the existence of finite spinorial representations for GL(4,$\mathbbm{C}$) which does not imply however the existence of finite spinorial representations for the double covering of GL(4,$\mathbbm{R}$).}. 
All quantities in this section should be considered as matrices, and we can omit the unit matrix $\mathbbm{1}$, so that e.g. $\eta_{ab}$ is understood as
$\eta_{ab}\mathbbm{1}$. Forms (except 0-forms) are denoted by bold symbols, e.g. $\bbe^a=\e^a{}_\mu\diff x^\mu$. Objects with spacetime indices are denoted by greek letters if they are connections (e.g. $\bLambda$) and by latin letters if they are tensors (e.g. $\bF$). 

Consider a finite transformation $\lambda$ generated with the infinitesimal parameters $\lambda^a{}_b$, in the case of the coframe $\bbe^a$,
\be
\bbe^a \rightarrow L^a{}_b(\lambda)\bbe^a\,, \quad L^a{}_b = \exp{\lp\frac{1}{2}\lambda^{cd}(\Delta^{(1)}_{cd}){}^a{}_b\rp}\,. \label{delta}
\ee
A spinor $\psi$ transforms according to a spinor representation 
\be
\psi \rightarrow L(\lambda)\psi\,, \quad  L = \exp{\lp \frac{1}{2}\lambda^{ab}\Delta^{(\frac{1}{2})}_{ab}\rp}\,.
\ee
At this point we do not assume anything about the transformation, so $L$ may stand for Lorentz as well as (General) Linear.
We also drop the argument $\lambda$ when it is unnecessary. 
Since the derivative of the spinor then transforms non-covariantly,
\be
\psi_{,\mu} \rightarrow L\psi_{,\mu} + L_{,\mu}\psi\,,
\ee
we introduce the covariant derivative $\Diff_\mu$ with the connection $\Gamma_\mu$ such that
\be
\Diff_\mu \psi = \psi_{,\mu} + \Gamma_\mu\psi\,, \quad \Gamma_\mu \rightarrow L\Gamma_\mu L^{-1}-L_{,\mu}L^{-1} \quad \Rightarrow \quad \Diff_\mu \psi \rightarrow L\Diff_\mu\psi\,.
\ee
Note that the matrix one-form $\bGamma$ is just an example of a gauge field $\bLambda$ such that for matrices with spacetime indices we can write $\Diff=\nabla + [\bGamma,\,\,\,]$.

The metric can be expressed in terms of the Dirac matrices $\gamma^\mu =  \gamma^a \ie_a{}^\mu$, as (note that in our convention $\{ \gamma^{\alpha},\gamma^{\beta}\} = 2\gamma^{(\alpha}\gamma^{\beta)}$)
\be \label{clifford}
\gamma^{(a}\gamma^{b)} = -\eta^{ab}\,, \quad \gamma^{(\mu}\gamma^{\nu)} = -g^{\mu\nu}\,.
\ee
The Hermitean property of Dirac matrices is $(\gamma^a)^\dagger = \gamma^0\gamma^a\gamma^0$. 
In the following we will make use of the identities which follow from the Clifford algebra (\ref{clifford})
\be  \label{clifford2}
\gamma^a\gamma^b\gamma^c = \eta^{ac}\gamma^b-2\eta^{b(a}\gamma^{c)} - i\epsilon^{dabc}\gamma_d\gamma^5 \quad
\Rightarrow \quad \lb \gamma^a\gamma^b,\gamma^c\rb = 4\eta^{c[a}\gamma^{b]}\,, \quad \left\{ \gamma^a\gamma^b,\gamma^c\right\} = 2\eta^{ab}+2i\epsilon^{dabc}\gamma_d\gamma^5\,,
\ee
where the $\gamma^5 = i\gamma^0\gamma^1\gamma^2\gamma^3\gamma^4$ is Hermitean, $(\gamma^5)^\dagger=\gamma^5$.  
The frame field $\bbie_a=\ie_a{}^\mu\partial_\mu$ is defined as the inverse of the coframe, $\bbe^b\cdot\bbie_a=\delta^a_b$, $\e^a{}_\nu\ie_a{}^\mu=\delta^\mu_\nu$. 
If we allow for non-metricity, $\Diff_\mu g^{\alpha\beta} = \nabla_\mu g^{\alpha\beta} = -Q_\mu{}^{\alpha\beta}$, the Dirac matrices can not be considered covariantly constant. If we start from the defining property of the Dirac matrices given in (\ref{clifford}) and perform an arbitrary variation of the metric  $\delta \eta^{ab}$ that induces a corresponding variation $\delta \gamma^a$ we obtain
\be
\delta\{\gamma^a,\gamma^b\}=\{\delta\gamma^a,\gamma^b\}+\{\gamma^a,\delta\gamma^b\}=-2\delta\eta^{ab},
\ee
whose general solution can be written as (see e.g. \cite{Treat1970,Weldon:2000fr})
\be
\delta \gamma^a=\frac12 \delta \eta^{ab}\gamma_b+[k,\gamma^a]
\ee
with $k\in \mathbbm{C}_{4\times4}$ arbitrary. This arbitrariness simply reflects the infinitesimal version with generator $k$ of the well-known fact that the Clifford algebra can be realised with equivalent sets of $\gamma$'s related by a similarity transformation. We can specify this general expression to the case when the variation in the metric corresponds to a covariant derivative so the equation reduces to
\be
\Diff\{\gamma^a,\gamma^b\}=\{\Diff\gamma^a,\gamma^b\}+\{\gamma^a,\Diff\gamma^b\}=2Q^{ab},
\ee
where we have used that $\Diff\eta^{ab}=-Q^{ab}$ (see \eqref{Qform} below), and the solution reads \cite{Treat1970,Weldon:2000fr} (see also \cite{Formiga:2012ns})
\be
\Diff\gamma^a=-\frac12 Q^a\,_b\gamma^b+[k,\gamma^a]\,.
\ee
The first piece in this expression is directly generated by the non-metricity and evinces the impossibility of having covariantly constant Dirac matrices in a non-metric space.  On the other hand, the arbitrariness encoded into $k$ remains even with vanishing non-metricity and reflects the non-triviality of the kernel of the covariant derivative of the Clifford algebra. As commented above, the non-trivial structure of the kernel is due to the freedom in performing a similarity transformation that preserves the Clifford algebra. We are thus free to choose a convenient representative among the equivalence class without affecting the physics and the usually adopted one consists in trivialising $k$ so that we have 
\begin{align} \label{presc}
\Diff \gamma^a =-\frac12 Q^a\,_b\gamma^b 
\qquad \text{or, equivalently} \qquad
\Diff_\mu \gamma^\alpha = -\frac{1}{2}Q_{\mu\nu}{}^\alpha \gamma^\nu\,.
\end{align}
The frame connection is related to the affine connection via
\be \label{framecon}
\Diff\bbe^a{} = 0 \quad \Rightarrow \quad \Lambda^a{}_{\mu b} = \e^a{}_\nu\lp \nabla_\mu\ie_b{}^\nu\rp = -\lp \nabla_\mu\e^a{}_\nu\rp\ie_b{}^\nu\,. 
\ee
By computing $\Diff\eta_{ab}$ we find the non-metricity one-form 
\be \label{Qform}
\bQ_{ab} = \diff\eta_{ab} - 2\bLambda_{(ab)}\,.
\ee
We shall adopt the orthonormal frame such that $\diff\eta_{ab}=0$. This implies that $\diff\gamma_a=0$ and $\Diff\gamma_a = -\bLambda_{(ab)}\gamma^b$, 
$\Diff\gamma^a = \bLambda^{(ab)}\gamma_b$. As shown in Appendix \ref{reps}, the spinor representation of the connection is given as
\be \label{bGamma}
\bGamma = -\frac{1}{4}\bLambda_{ab}\gamma^a\gamma^b - \frac{1}{8}\bZ\,,
\ee
where $\bZ$ is an arbitrary one-form.
Let us do a consistency check by computing (\ref{presc}):
\ba
\Diff_\mu\gamma^\alpha & = & \nabla_\mu\lp \gamma^c\ie_c{}^\alpha\rp - \frac{1}{4}\Lambda_{a\mu b}\lb \gamma^a\gamma^b,\gamma^c\rb \ie_c{}^\alpha\nonumber \\
& = & \gamma^c \Lambda^a{}_{\mu c}\ie_a{}^\alpha - \Lambda_{a\mu b}\eta^{c[a}\gamma^{b]}\ie_c{}^\alpha \nonumber \\
& = & \gamma_a \Lambda^{(a}{}_\mu{}^{b)}\ie_b{}^\gamma  = -\frac{1}{2}Q_{\mu\nu}{}^\alpha \gamma^\nu\,.
\ea
In the first line we have only used the definitions of the spacetime Dirac matrices and the covariant derivative, in the second line the relation (\ref{framecon})
and the identity (\ref{clifford2}), and in the third line recalled (\ref{Qform}) in the orthonormal frame. 

Consider the Hermitean Dirac action for a spinor $\psi$ with mass $m$ 
\be \label{dirac1}
I_\psi = -\frac{1}{2}\int\diff^4 x  \sqrt{-g}\lb \lp i\bar{\psi}\gamma^\mu \Diff_\mu\psi\rp + \lp i\bar{\psi}\gamma^\mu \Diff_\mu\psi\rp^\dagger - 2 m\bar{\psi}\psi\rb\,,
\ee
where $\bar{\psi}=\psi^\dagger\gamma^0$ is the conjugate spinor. More explicitly, we have
\be
I_\psi = -\int\diff^4 x  \sqrt{-g}\lb \frac{i}{2}\lp\bar{\psi} \gamma^\mu\partial_\mu\psi - \partial_\mu\bar{\psi}\gamma^\mu\psi  \rp+ \bar{\psi}\lp i\Gamma^{\text{H}} - m\rp  \psi\rb\,, \quad
\text{where} \quad \Gamma^{\text{H}} = \frac{1}{2}\lp\bgamma\cdot\bGamma - \gamma^0 \bGamma^\dagger\gamma^0\cdot\bgamma\rp\,. 
\ee
We have denoted the vector $\bgamma=\gamma^\mu\partial_\mu$.
Plugging in (\ref{bGamma}) gives
\be
\Gamma^{\text{H}} =- \frac{1}{8}\text{Re}(\bLambda^{ab})\cdot\{\bgamma,\gamma_{[a}\gamma_{b]}\} - 
\frac{i}{8}\text{Im}(\bLambda^{ab})\cdot[ \bgamma,\gamma_{[a}\gamma_{b]}]  -\frac{i}{8}\text{Im}(\bQ+\bZ) \cdot\bgamma\,,
\ee
which becomes, by using (\ref{clifford2}),
\be \label{gamma2}
\Gamma^{\text{H}} =  - \frac{i}{4}\epsilon^{abcd}\text{Re}(\bLambda_{ab})\cdot\bbie_c\gamma_d\gamma^5
+ \frac{i}{2} \text{Im}(\bLambda^{[ab]})\cdot\bbie_a\gamma_b -\frac{i}{8}\text{Im}(\bQ+\bZ) \cdot\bgamma\,.
\ee
Thus, instead of being coupled to $\bgamma\cdot\bGamma$, the spinor is coupled to $\Gamma^{\text{H}}$.

At this point, it is useful to recall the well-known decomposition of the GL(4) connection,
\be \label{decomposition}
A^{a}{}_{bc} = \omega^{a}{}_{bc} + K^{a}{}_{bc} + L^{a}{}_{bc}\,,
\ee
where the Levi-Civita connection $\omega^{a}{}_{bc}$, the contortion tensor $K^a{}_{bc}$ and disformation tensor $L^a{}_{bc}$,
\be 
\omega^a{}_{bc} = \frac{1}{2}\Omega^a{}_{bc} - \Omega_{(bc)}{}^{a}\,, \quad  K^a{}_{bc} = \frac{1}{2}T^a{}_{bc} - T_{(bc)}{}^{a}\,, \quad L^a{}_{bc} = \frac{1}{2}Q^a{}_{bc} - Q_{(bc)}{}^{a}\,,
\ee
are given by the coefficients of anholonomy $\Omega_{abc}$,
\be \label{anholonomy}
\diff\bbe^a = -\frac{1}{2}\Omega^a{}_{bc}\bbe^b\wedge\bbe^c \quad \Leftrightarrow \quad \Omega^a{}_{bc} = \bbe^a\cdot\lb \bbie_b,\bbie_c\rb
\quad \Leftrightarrow \quad \Omega^a{}_{bc} = 2\e^a{}_{[\mu,\nu]}\ie_b{}^\mu\ie_c{}^\nu\,,
\ee
the torsion $T^a{}_{bc} = \Diff\bbe^a\cdot\bbie_b\cdot\bbie_c$ and the nonmetricity $Q_{abc}=\Diff\eta_{bc}\cdot\bbie_a$ of the connection, respectively. 
There are six independent objects one can obtain by different contractions of the components of the connection,
\be
\omega_a = \omega^b{}_{ba}\,, \quad
\tilde{\omega}_a = \epsilon_{abcd}\omega^{bcd}\,, \quad
T_a = T^b{}_{ab}\,, \quad
\tilde{T}_ a =  \epsilon_{abcd}T^{bcd}\,, \quad 
Q_a = Q_{ab}{}^b\,, \quad
\tilde{Q}_a = Q_{ba}{}^b\,.
\ee
The pieces relevant to the Hermitean version of the spin connection (\ref{gamma2}) are given by 
\ba
\epsilon^{abcd}\bA_{ab}\cdot\ie_c & = & \epsilon^{abcd}A_{acb} =  -\tilde{\omega}^d - \tilde{T}^d\,, \\
\bA^{[ab]}\cdot\ie_a & = & A^{[a}{}_{a}{}^{b]} = \omega^b - T^b - \frac{1}{2}\lp Q^b-\tilde{Q}^b\rp\,.  
\ea 
We can then decompose (\ref{gamma2}) as follows:
\be
i\Gamma^\text{H} = i\gamma^\mu\Gamma^\text{H}_\mu = \gamma^\mu\lp \gamma^5\Phi_\mu + \Psi_\mu\rp\,,
\ee
wherein the real and the imaginary parts of the affine connection enter as
\begin{subequations}
\ba
\bPhi & = &  \frac{1}{4}\text{Re}\lp\tilde{\bomega} + \tilde{\bT}\rp\,, \\
\bPsi & = &  - \frac{1}{2}\text{Im}\lp \bomega - \bT + \frac{1}{2}\tilde{\bQ}\rp +  \frac{1}{8}\text{Im}\lp 3\bQ + \bZ\rp\,, \label{imag}
\ea
\end{subequations}
respectively.
This verifies the claim of the Lemma {\bf 2}. To wit, if the connection is real, $\bPsi=0$, and devoid of axial torsion, $\tilde{\bT}=0$, only the Levi-Civita part contributes to the action (\ref{dirac1}) through $\tilde{\bomega}$. This property was used to show the viability of certain vector distorted geometries in \cite{Jimenez:2015fva}.


\section{Implications for fermions}
\label{spinor2}

Let $I=I_G(g,\nabla)+I_\phi(g,\nabla,\phi)$ be an action for a coupled matter-gravity system. The variations of $I$ w.r.t. the geometric variables define the metric field equations, the connection excitation and the hypermomentum as
\be
\mathcal{E}_{\mu\nu} =  \frac{\delta I}{\delta g^{\mu\nu}}\,, \quad 
\mathcal{P}^{\mu\nu}{}_\alpha = \frac{\delta I_G}{\delta \Gamma^\alpha{}_{\mu\nu}}\,, \quad 
\mathcal{H}^{\mu\nu}{}_\alpha = -\frac{\delta I_\phi}{\delta \Gamma^\alpha{}_{\mu\nu}}\,,
\ee
respectively. For this generic action considered in the geometrical setting with arbitrary $\Gamma$, holds the following
\newline
\newline
\underline{{\bf Lemma 4.}}  The {\it generalised Noether identity} resulting from the diffeomorphism invariance of $I$ is
\be
\mathcal{D}_\mu \mathcal{E}^{\mu}{}_\nu = \Big[ \delta^\rho_\nu\lp  \nabla_\alpha\nabla_\beta+ 4T_{(\alpha}\nabla_{\beta)} + 2\nabla_\beta T_\alpha  +  T_\alpha T_\beta \rp -2T^\rho{}_{\mu\beta}\nabla_\alpha  - 4T_\alpha T^\rho{}_{\nu\beta} - R^\rho{}_{\mu\alpha\beta}  \Big]\lp \mathcal{P}^{\alpha\beta}{}_\rho -\mathcal{H}^{\alpha\beta}{}_\rho\rp + \Diff^{(\phi)}_\nu\left( \frac{\delta I_\phi}{\delta \phi}\cdot\phi\right)\,,\nn 
\ee
with $\Diff_\mu^{(\phi)}\phi$ some derivative that depends on the type of matter field and $\cdot$ stands for a sum over internal indices.\\
{\bf Proof} : See Appendix \ref{noether}.  This gives an explicit form for the generalised Bianchi identity \cite{Koivisto:2005yk}, which is useful in applications to particular geometries (one may consider the lagrange multipliers that impose the desired geometry to be included in $\phi$).  In the three special cases we will consider below, the connection equation of motion we state could also be easily deduced from the derivations of \cite{BeltranJimenez:2018vdo}. 

We shall now specialise to the case of a fermion field $I_\phi=I_\psi$. Separating contributions from the real and the possible imaginary parts of the connection,
\be \label{hyper}
\text{real:} \quad
\mathcal{H}^{\mu\nu}_\psi{}_\alpha = -\frac{1}{4}\sqrt{-g}g_{\rho\alpha}\epsilon^{\rho\mu\nu\beta}\bar{\psi}\gamma_\beta\gamma^5\psi\,,
\quad
\text{imaginary:} \quad
\mathcal{H}^{\mu\nu}_\psi{}_\alpha = -\frac{1}{4}\sqrt{-g}\delta^\mu_\alpha\gamma^\nu\,.
\ee
We can then consider different cases of interest.
\begin{itemize}

\item {\bf Palatini theory}. The connection equation of motion is $\mathcal{P}^{\mu\nu}{}_\alpha = \mathcal{H}^{\mu\nu}{}_\alpha$. In the case $I_G \sim \int \diff^4 x \sqrt{-g} R$ corresponding to case of Einstein-Cartan-Kibble-Sciama theory coupled to spinors, we obtain $\sqrt{-g}(T^{\mu}{}_\alpha{}^\nu + \delta^\mu_\alpha T^\nu - T_\alpha g^{\mu\nu} + Q_{\alpha}{}^{\mu\nu} - \delta^\mu_\alpha\tilde{Q}^\nu + Q^{[\mu}\delta^{\nu]}_\alpha) \sim \mathcal{H}^{\mu\nu}_\psi{}_\alpha$, which is solved by a metric-compatible connection with axial torsion proportional to (\ref{hyper}). As it is well-known, this results in a four-fermion contact interaction which only becomes relevant at extreme densities \cite{Hehl1976GeneralRW}.

\item {\bf General} (including metric) {\bf teleparallelism}. The connection equation of motion is $(\nabla_\mu+T_\mu)\mathcal{P}^{\mu[\nu\alpha]} = (\nabla_\mu+T_\mu)\mathcal{H}^{\mu\nu\alpha}$. 
In the case of the teleparallel equivalent of General Relativity, the left hand side vanishes identically, resulting in an additional constraint for spinors in the presence of torsion. At the Minkowski limit, the constraint is degenerate with the conservation law $\partial_\mu j^\mu=0$ derived below, but in a generic gravitational system probably leads to an inconsistency, as has been claimed previously.

\item {\bf Symmetric teleparallelism}. The connection equation of motion is $\nabla_\mu\nabla_\nu\mathcal{P}^{\mu\nu}{}_\alpha = \nabla_\mu\nabla_\nu\mathcal{H}^{\mu\nu}{}_\alpha$.
Because of the antisymmetry of (\ref{hyper}) and the commutative property of the symmetric teleparallel covariant derivative, the right hand side vanishes identically for the
real part in (\ref{hyper}). The contribution from the possible imaginary part is guaranteed to vanish due to the conservation of the probability current. 
Thus, the
hypermomentum of spinors is irrelevant to the dynamics of gravitation. In the case of Coincident General Relativity, also the left hand side vanishes identically. 
\end{itemize}
For completeness, the energy-momentum tensor of spinors is given as 
\be
\frac{1}{\sqrt{-g}}\frac{\delta I_\psi}{\delta g_{\mu\nu}} = -\frac{i}{2}\lb\bar{\psi}\gamma^\alpha\nabla^\text{H}_\alpha\psi-\lp\nabla^\text{H}_\alpha\bar{\psi}\rp\gamma^\alpha\psi\rb g^{\mu\nu}
+\frac{i}{2}\lb\bar{\psi}g^{\alpha(\mu}\gamma^{\nu)}\nabla^\text{H}_\alpha\psi - g^{\alpha(\mu}\gamma^{\nu)}\lp \nabla^\text{H}_\alpha\bar{\psi}\rp\psi\rb + m\bar{\psi}\psi\,,
\ee
and the equations of motion $\delta I_\psi/ \delta \bar{\psi} = \delta I_\psi/ \delta\psi =0$ are  
\begin{subequations}
\ba
i\gamma^\mu\lp\partial_\mu+ \Gamma^\text{H}_\mu\rp \psi +\frac{i}{2\sqrt{-g}}\partial_\mu\lp\sqrt{-g}\gamma^\mu\rp\psi -m\psi & = &  0\,, \\
i\lp \partial_\mu\bar{\psi} - \Gamma^\text{H}_\mu\bar{\psi}\rp\gamma^\mu  +\frac{i}{2\sqrt{-g}}\bar{\psi}\partial_\mu\lp\sqrt{-g}\gamma^\mu\rp +\bar{\psi}m & = &  0\,.
\ea 
\end{subequations}
Using the formulae (\ref{anholonomy}) and the constancy $\diff \gamma^a=0$ of the Dirac matrices, we can alternatively write
\begin{subequations}
\ba
\lp i\gamma^\mu \partial_\mu+ \gamma^\mu \Gamma^\text{H}_\mu  + \frac{i}{2} \gamma^\mu\omega_{\mu} -m\rp \psi & = &  0\,, \\
\lp \partial_\mu\bar{\psi}\rp i\gamma^\mu - \bar{\psi}\lp i\gamma^\mu\Gamma^\text{H}_\mu  - \frac{i}{2} \gamma^\mu\omega_{\mu} - m\rp & = &  0\,. 
\ea 
\end{subequations}
It is easy to see that the probability current, $j^\mu = \sqrt{-g}\bar{\psi}\gamma^\mu\psi$, is conserved, $\partial_\mu j^\mu = 0$. 

It would seem very challenging to experimentally constrain the precise form of the coupling of spinors to the gravitational connection. 
We can obtain a second order evolution equation for the projected one-component spinor 
\be \label{onec}
\phi = \frac{1}{2}\lp 1 + \gamma^5\rp\psi\,. 
\ee
Let us define a short-hand notation and restore the Planck constant,
\be
\hat{\nabla}_\mu\psi = \lb\partial_\mu+ \Gamma^\text{H}_\mu + \frac{1}{2}\omega_\mu \rb\psi \quad \text{i.e.} \quad
\lp i\hbar {\gamma}^\mu\hat{\nabla}_\mu -m\rp\psi = 0\,.
\ee
Computing now $\lp i {\bgamma}\cdot\hat{\nabla} -m\rp\phi$ from (\ref{onec}) and noting that $\Gamma^{\text{H}}\gamma^5=-\gamma^5\Gamma^{\text{H}}$ we obtain the desired second-order equation
\be \label{2order}
\lp i\hbar {\gamma}^\mu\hat{\nabla}_\mu -m\rp\lp i\hbar {\gamma}^\mu\hat{\nabla}_\mu + m\rp\phi = 0\,.
\ee
In the semi-classical approximation one may consider the Ansatz $\phi = \exp{(i S/\hbar)}\phi_0$, where $S$ is very large in units of $\hbar$.  
Then (\ref{2order}) reduces to
\be
g^{\mu\nu}S_{,\mu}S_{,\nu} + m^2 = \hbar\left[ \gamma^\alpha\partial_\alpha\lp \gamma^\mu S_{,\mu}\rp - g^{\mu\nu}\hat{\Gamma}_\mu S_{,\nu}\right] - 
\hbar^2\left[
\gamma^\alpha\partial_\alpha\lp \gamma^\mu \hat{\Gamma}_\mu\rp +  g^{\mu\nu}  \hat{\Gamma}_\mu \hat{\Gamma}_\nu\right]\,.
\ee 
At the leading order this describes the dispersion relation $g^{\mu\nu}k_\mu k_\nu = -m^2$, and the trajectories become the metric geodesics.
Only a correction proportional to the $\hbar$ appears to the above equation which is dependent on the independent connection.  A modified dispersion relation at the lowest order could occur on a non-trivial background configuration for the connection.

 
\section{Conclusions and discussion}
\label{concs}

Complementary perspectives to gravity emerge from different geometrical formulations, wherein one may interpret a given theory in terms of curvature, torsion, or non-metricity. An instance of this is the ternion of geometrical representations of General Relativity. Nevertheless, subtleties and ambiguities might arise in generalised geometries when matter couplings have to be considered as well. If one starts with the usual point particle $I$ that extremises the purely metrical quantity, the proper time, one obtains the equation motion in terms of solely the Levi-Civita connection, wrt which the autoparallels coincide with the geodesics. This intuitive result is also the natural consequence of MCP
for bosonic and fermionic fields in spacetimes equipped with only the metric connection. For general spacetimes MCP does not necessarily give rise to the same standard matter coupling, especially if torsion is present.

In this paper we investigated the coupling of the standard matter and gauge fields to spacetime geometry, leaving the detailed study of non-canonical scalar, vector and other fields elsewhere. Then, from the Lemmas {\bf 1} and {\bf 3} now follow the 
\newline
\newline
{{\bf Conclusion a)}}:  Spacetime torsion, if it exists, is non-minimally coupled.
\newline
\newline
That the spacetime torsion has to couple to matter in some non-minimal manner, e.g. according to MCP, is required generically to save the gauge symmetries of the standard model. In teleparallel models particularly, it is in addition required for consistent dynamics of elementary particles.  -- In the symmetric affine sector, the Lemmas {\bf 1}-{\bf 3} justify the
\newline\newline
{\bf Conclusion b)}: $\bR_{(ab)}$ measures the 2$^{\text{nd}}$ clock effect. In a torsion-free spacetime MCP=mCP.  
\newline
\newline
As it should be clear from our results, the second clock effect is absent for physical particles. Therefore, minimally coupled gauge and matter fields interact with an arbitrary symmetric and non-metric connection - even the vanishing connection of Coincident General Relativity - exactly as they do with the metric-compatible Levi-Civita connection. However, the non-metric i.e. symmetric curvature is a gauge-invariant measure of the path-dependent discalibration, whereas the more familiar metric i.e. Riemann curvature is a measure of the local rotation.   

It should be clarified that by spacetime geometry we mean the real components of both the metric and the affine structures. This excludes beyond the scope of the present paper the possible relation between the imaginary components of the affine connection and the gauge fields of internal interactions\footnote{In the context of ``purified gravity'', it has been speculated that the continuum of real numbers spans the integrable quotient $M$, and the rest is a computation in the three other division algebras \cite{Koivisto:2018aip,Koivisto:2019ejt}. The integrability (and a fortiori, teleparallelism) of classical gravity is due to that the Planck mass is the mass of the gravitational field $\Gamma$ \cite{Koivisto:2019jra}.}. 

Needless to say, one need not to follow MCP, but can regard it only as a procedure that works in some theories but ought not to be naively extrapolated to others. In any case, MCP enforces some technical economy, and reflects both the logic of gauge theory and the unique, universal character of gravity expressed in the equivalence principle. It yet remains to be investigated how much further MCP may guide us. 

Embracing MCP is understanding the limitations it puts to a theory. A well defined theory must have a {\it canonical} choice of the generalisation of the spacetime derivatives in {\it canonical} inertial coordinates; the generalisation must be uniquely determined for any representation and for any geometrical construction including objects such as frames, volumes, determinants, products such as dual, star, wedge, derivatives such as exterior, adjoint, Lie, etc which are all available if a suitable manifold structure is postulated; the generalisation $\nabla$ is not arbitrary, but determined by G, or even more properly, by $I(\eta,\phi,\partial\phi)$ which encodes both the fields $\phi$, their symmetry G, and the necessary details included in the action formulation $I$ which typically amounts to the instructions for integrating the quotient of G that is interpreted as the spacetime manifold; and if still carried further, the principle should dictate also the dynamics of the gravitational fields, and would then for example exclude the case $I_{EH}(\phi,\partial\phi,\partial^2 \phi)=\int g(\mathcal{D}^2)$ known as the Einstein-Hilbert action whilst allowing to take terms such as $\int g(g,\nabla g)$ into consideration.      

We end with a proposition. It is based upon a curious feature of MCP, $I_\phi(g,\phi,\mathcal{D}\phi)=I_\phi(g,\phi,\partial\phi)$, that is already there in standard General Relativity (but not in its Palatini variant) and upon (the first part of) the Conclusion {\bf b)} that carries the feature into the symmetric teleparallel spacetime so that also there $I_\phi(g,\phi,\nabla\phi)=I_\phi(g,\phi,\partial\phi)$. 
\newline
\newline
\underline{{\bf Proposition 1.}} Given the physics $I_\phi(\eta,\phi,\partial\phi)$ in an inertial frame, its classical gravity is determined\footnote{The integral $I_E$ is appropriately called as the Einstein action \cite{1916SPAW......1111E}. The unique  
form of the action $I_G=\int \diff^4 x\mathcal{Q}$ from which the pure translation $\Gamma$ decouples \cite{BeltranJimenez:2017tkd} can be deduced from elementary classical physics principles \cite{Koivisto:2019jra}.} by  
\be
\text{\it the minimal decoupling principle } \quad : \quad   
\underbrace{I_G(\eta,\partial\eta)}_{=0} + \underbrace{I_\phi(\eta,\phi,\partial\phi)}_{\text{$@$inertial frame}} \rightarrow \underbrace{I_G(g,\nabla g)}_{=\int \diff^4 x\mathcal{Q} \newline \\ \mathring{=} I_E(g,\partial g)} + \underbrace{I_\phi(g,\phi,\nabla\phi)}_{=I_\phi(g,\phi,\partial\phi)}\,. \nn \quad\quad \text{(MDP)}
\ee
A certain coincidence occurs also in the $\phi$-sector. 
\newline
\newline
{\bf Acknowledgements}:  JBJ acknowledges support from the  {\it Atracci\'on del Talento Cient\'ifico en Salamanca} programme and the MINECO's projects FIS2014-52837-P and FIS2016-78859-P (AEI/FEDER). LH is supported by funding from the European Research Council (ERC) under the European Unions Horizon 2020 research and innovation programme grant agreement No 801781 and by the Swiss National Science Foundation grant 179740. TSK was funded by the Estonian Research Council PRF project PRG356 and by the European Regional Development
Fund CoE TK133. This article is based upon work from CANTATA COST (European Cooperation in Science and Technology) action CA15117, EU Framework Programme Horizon 2020.


\appendix

\section{Generalized Noether identity}
\label{noether}
In this Appendix we will derive the Noether identities corresponding to the gauge diffeormophisms invariance of the action. These results can be found in the literature (see e.g. \cite{Hehl:1994ue,Iosifidis:2020gth}), but we will reproduce them here for completeness. We will also seize the opportunity to clarify some potentially confusing points. Let us then consider the following action
\begin{equation}
I= I_{G}[g,\Gamma]+\frac{2}{M_P^2} I_{\phi}[g,\Gamma,\phi]\,, \label{maglag}
\end{equation}
where $g$ and $\Gamma$ represent the gravity sector and $\phi$ stands for the matter fields. We will work with the spacetime version of the theory, i.e., all quantities in the general linear bundle are translated to the spacetime tangent bundle. The variations of the gravitational fields under a diffeomorphism generated by the vector field $v^\mu$ are given by the Lie derivatives \cite{schouten1954ricci}
\begin{eqnarray}
\delta_vg_{\mu\nu}&=&-\Lie_vg_{\mu\nu}=-\Big(v^\lambda \partial_\lambda g_{\mu\nu}+2g_{\lambda(\mu}\partial_{\nu)} v^{\lambda}\Big),\\
\delta_v\Gamma^\alpha{}_{\mu\nu}&=&-\Lie_v\Gamma^\alpha{}_{\mu\nu}=-\Big(v^\lambda\partial_\lambda\Gamma^\alpha{}_{\mu\nu}-\partial_\lambda v^\alpha\Gamma^\alpha{}_{\mu\nu}+\partial_\mu v^\lambda\Gamma^\alpha{}_{\lambda\nu}+\partial_\nu v^\lambda\Gamma^\alpha{}_{\mu\lambda}+\partial_\mu\partial_\nu v^\alpha\Big).
\end{eqnarray}
These variations can be expressed in terms of manifestly tensorial quantities and the affine connection as follows:
\begin{eqnarray}
\delta_v g_{\mu\nu}&=&-2g_{\lambda(\mu}\nabla_{\nu)}v^\lambda+\Big(2 T_{(\mu\nu)\lambda}-Q_{\lambda\mu\nu}\Big)v^\lambda \,,\\
\delta_v\Gamma^\alpha{}_{\mu\nu}&=&-\nabla_\mu\nabla_\nu v^\alpha+T^\alpha{}_{\nu\lambda}\nabla_\mu v^\lambda+\Big(R^\alpha{}_{\nu\mu\lambda}+\nabla_\mu T^\alpha{}_{\nu\lambda}\Big)v^\lambda \,.
\label{eq:Liemetricconnection2}
\end{eqnarray}
The transformation for the metric can be substantially simplified by recalling the usual GR result that is of course also in the general affinely connected space
\begin{equation}
\delta_v g_{\mu\nu}=-2g_{\lambda(\mu}{\mathcal{D}}_{\nu)}v^\lambda=-2{\mathcal{D}}_{(\mu}v_{\nu)}\,,
\label{eq:Liemetric2}
\end{equation}
so the metric changes with its Levi-Civita connection. There is nothing special about the metric in this sense and this transformation law applies to any symmetric rank-2 tensor. We can then express the variation of the action as
\begin{equation}
\delta_v I=-\int\diff^4 x\frac{\delta  I}{\delta\Psi^A}\Lie_v\Psi^A
\end{equation}
with $\Psi^A=\{g_{\mu\nu}, \Gamma^\alpha{}_{\mu\nu},\cdots\}$, where the dots stand for all other possible matter fields. As a proxy of the matter sector for illustrative purposes and for the sake of simplicity we will take a set of (Diffs-)scalar fields $\varphi^a$,which could nevertheless belong to some non-trivial representation of an internal group. They transform as
\begin{equation}
\delta_v\varphi^a=-\Lie_v\varphi^a=-v^\lambda\partial_\lambda\varphi^a\,.
\end{equation}
The extension to other fields is straightforward and it is just necessary to include the non-trivial dragging terms in the Lie derivative. If we introduce the functional variations
\begin{equation}
\mE^{\mu\nu}\equiv\frac{\delta I}{\delta g_{\mu\nu}},\quad \mF_\alpha{}^{\mu\nu}\equiv\frac{\delta I}{\delta \Gamma^\alpha{}_{\mu\nu}}\quad\text{and}\quad\mE_a\equiv\frac{\delta I}{\delta \varphi^a}
\end{equation}
that give the corresponding field equations, we find
\begin{eqnarray}
\delta_v I&=&\int\diff^4x\left[2\nablat_\mu\mE^\mu{}_\lambda+\Big(2T_{\mu\nu\lambda}-Q_{\lambda\mu\nu}\Big)\mE^{\mu\nu}
-\nablat_\nu\nablat_\mu \mF_\lambda{}^{\mu\nu}+T^\alpha{}_{\lambda\nu}\nablat_\mu \mF_\alpha{}^{\mu\nu}+\Big(R^\alpha{}_{\nu\mu\lambda}+T^{\beta}{}_{\beta\mu} T^\alpha{}_{\nu\lambda}\Big)\mF_\alpha{}^{\mu\nu}-\mE_a\partial_\lambda\varphi^a\right]v^\lambda\nonumber\\
&&+\int\diff^4x\,\partial_\mu\Big(\mathcal{J}^\mu{}_{\lambda}v^\lambda+\mathcal{J}^{\mu\nu}{}_\lambda\partial_\nu v^\lambda\Big)\,,
\end{eqnarray}
where we have defined the derivative $\nablat_\mu\equiv\nabla_\mu-T^\alpha{}_{\alpha\mu}$ that arises from integration by parts and $\mathcal{J}^\mu{}_{\lambda}$ and $\mathcal{J}^{\mu\nu}{}_\lambda$ are two density currents that depend on the different fields. The second term originates from the inhomogeneous piece of the connection transformation that depends on second derivatives of the gauge parameter. This is a distinctive feature of metric-affine theories that does not appear in e.g. Yang-Mills theories. In order to obtain the off-shell conserved currents we need to impose that both the gauge parameters and their derivatives vanish on the boundary\footnote{The additional condition on the behaviour of the gauge parameter on the boundary may have interesting non-trivial consequences for the infrared structure of the theories, asymptotic charges of spacetimes with boundaries.} so that we obtain the set of identities:
\begin{equation}
2\nablat_\mu\mE^\mu{}_\lambda+\Big(2T_{\mu\nu\lambda}-Q_{\lambda\mu\nu}\Big)\mE^{\mu\nu}
-\nablat_\nu\nablat_\mu \mF_\lambda{}^{\mu\nu}+T^\alpha{}_{\lambda\nu}\nablat_\mu \mF_\alpha{}^{\mu\nu}+\Big(R^\alpha{}_{\nu\mu\lambda}+T^{\beta}{}_{\beta\mu} T^\alpha{}_{\nu\lambda}\Big)\mF_\alpha{}^{\mu\nu}=\mE_a\partial_\lambda\varphi^a.
\end{equation}
These are the general identities derived from the diffeomorphism-invariance of the action. The three pieces coming from the metric conspire to give $2{\mathcal{D}}_\mu\mE^\mu{}_\lambda$ so the identities can alternatively be written as
\begin{equation}
2\mathcal{D}_\mu\mE^\mu{}_\lambda
-\nablat_\nu\nablat_\mu \mF_\lambda{}^{\mu\nu}+T^\alpha{}_{\lambda\nu}\nablat_\mu \mF_\alpha{}^{\mu\nu}+\Big(R^\alpha{}_{\nu\mu\lambda}+T^{\beta}{}_{\beta\mu} T^\alpha{}_{\nu\lambda}\Big)\mF_\alpha{}^{\mu\nu}=\mE_a\partial_\lambda\varphi^a.
\end{equation}
Of course, there is no secret conspiracy and this is the form of the identities we would have obtained had we used \eqref{eq:Liemetric2} instead of \eqref{eq:Liemetricconnection2}. For a general matter field, the right hand side of the Bianchi identities would be given by some differential operator. If instead of a scalar we considered an arbitrary $(p,q)$-tensor ${\mathcal A}^{\mu_1\cdots\mu_p}{}_{\nu_1\cdots\nu_q}$, possibly with some internal indices as well, from the transformation rule $\delta_v{\mathcal A}=-\Lie_v{\mathcal A}$, the right hand side of the Bianchi identities would be given by the following covariant derivative:
\begin{align}
&\Diff_\lambda^{(\mathcal A)}\left[\frac{\delta I}{\delta {\mathcal A}^{\mu_1\cdots\mu_p}{}_{\nu_1\cdots\nu_q}}{\mathcal A}^{\mu_1\cdots\mu_p}{}_{\nu_1\cdots\nu_q}\right]\equiv\frac{\delta I}{\delta {\mathcal A}^{\mu_1\cdots\mu_p}{}_{\nu_1\cdots\nu_q}}\partial_\lambda{\mathcal A}^{\mu_1\cdots\mu_p}{}_{\nu_1\cdots\nu_q}\nonumber\\
&-\partial_{\nu_1}\left(\frac{\delta I}{\delta {\mathcal A}^{\mu_1\cdots\mu_p}{}_{\nu_1\cdots\nu_q}}{\mathcal A}^{\mu_1\cdots\mu_p}{}_{\lambda\nu_2\cdots\nu_q}\right)-\cdots-\partial_{\nu_q}\left(\frac{\delta I}{\delta {\mathcal A}^{\mu_1\cdots\mu_p}{}_{\nu_1\cdots\nu_q}}{\mathcal A}^{\mu_1\cdots\mu_p}{}_{\nu_1\cdots\nu_{q-1}\lambda}\right)\nonumber\\
&+\partial_{\kappa}\left(\frac{\delta I}{\delta {\mathcal A}^{\mu_1\cdots\mu_p}{}_{\nu_1\cdots\nu_q}}{\mathcal A}^{\kappa\mu_2\cdots\mu_p}{}_{\nu_1\cdots\nu_q}\right)\delta^{\mu_1}{}_\lambda+\cdots+\partial_{\kappa}\left(\frac{\delta I}{\delta {\mathcal A}^{\mu_1\cdots\mu_p}{}_{\nu_1\cdots\nu_q}}{\mathcal A}^{\mu_1\cdots\mu_{p-1}\kappa}{}_{\nu_1\cdots\nu_q}\right)\delta^{\mu_p}{}_\lambda ,
\end{align}
where a sum over internal indices is implied. Notice that the covariant character of this expression is guaranteed by its own definition even though it is not manifestly covariant. Furthermore, fields that do not transform as proper tensors under diffeomorphisms (e.g. spinors or tensorial densities) will feature a different expression for this covariant derivative but it will always be determined by the corresponding Lie derivative. If we restrict to the pure gravity sector, i.e. the sector of the action that does not depend on the matter fields, the right hand side of the Bianchi identities vanishes and we obtain 
\begin{equation}
2\mathcal{D}_\mu\mathcal{G}^\mu{}_\lambda
-\nablat_\nu\nablat_\mu  \mathcal{P}_\lambda{}^{\mu\nu}+T^\alpha{}_{\lambda\nu}\nablat_\mu  \mathcal{P}_\alpha{}^{\mu\nu}+\Big(R^\alpha{}_{\nu\mu\lambda}+T^{\beta}{}_{\beta\mu} T^\alpha{}_{\nu\lambda}\Big) \mathcal{P}_\alpha{}^{\mu\nu}=0
\end{equation}
with
\be
\mathcal{G}^{\mu\nu}\equiv\frac{\delta I_G}{\delta g_{\mu\nu}} \quad \text{and}\quad
\mathcal{P}^{\mu\nu}{}_\alpha \equiv \frac{\delta I_G}{\delta \Gamma^\alpha{}_{\mu\nu}}.
\ee
It may be convenient to stress that for these identities to hold, it is not necessary that the matter fields are on-shell. This simply follows from imposing diffemorphism invariance for the piece $ I_G$ in \eqref{maglag}. In a pure metrical theory without any dependence on the connection, this equation recovers the standard Bianchi identities $\mathcal{D}_\mu\mathcal{G}^\mu{}_\lambda=0$. It is also important to realise that these metric Bianchi identities will be satisfied in a general metric-affine theory in any sector that is decoupled from the connection. That is for instance the case of the bosonic sector of the theory with minimal couplings.


\section{On representations}
\label{reps}

The cotangent space of the GL(4) group can be spanned by the 16 vectors $\Sigma_{ab}=2x_a\partial_b$, with the commutation relations
\be
[\Sigma_{ab}, \Sigma_{cd}] = 2\lp \eta_{bc}\Sigma_{ad} - \eta_{ad}\Sigma_{cd}\rp\,.
\ee 
The Killing vectors can be splitted into the Lorentz rotations $r_{ab}=\Sigma_{[ab]}$ and the shear generators $q_{ab}=\Sigma_{(ab)}$, with the algebra
\be
[r_{ab},r_{cd}] = 2\lp \eta_{d[a}r_{b]c} - \eta_{c[a}r_{b]d}\rp\,, \quad
[r_{ab},q_{cd}] =-2\lp \eta_{d[a}q_{b]c} + \eta_{c[a}q_{b]d}\rp\,, \quad
[q_{ab},q_{cd}] = 2\lp \eta_{d(a}r_{b)c} + \eta_{c(a}r_{b)d}\rp\,.
\ee
The infinitesimal gauge transformations are given by the Lie derivatives along the Killing vectors. For example, for the transformation of a vector $\bV$, we get
\be
\mathcal{L}_{\Sigma_{ab}}\bV = [\Sigma_{ab}, \bV] = \lp \Sigma_{ab}V^c - 2\eta_{d[a}\delta^c_{b]}V^d\rp\partial_c = \lb (\Sigma_{ab}^{(0)})\delta^c_d + (\Delta_{ab}^{(1)})^c{}_d\rb V^d\partial_c\,,
\ee
where the second piece, the matrix part of the operator (the first piece being called the orbital part of the operator acting upon $\bV$) defines the 
vector representation we referred to in (\ref{delta}), 
\be
(\Delta_{ab}^{(1)})^c{}_d = -2\eta_{da}\delta^c_b\,, \quad \text{i.e.} \quad (r_{ab}^{(1)})^c{}_d = r_{ab}\delta^c_d-2\eta_{d[a}\delta^{c}_{b]} \quad \text{and} \quad
 (q_{ab}^{(1)})^c{}_d = q_{ab}\delta^c_d-2\eta_{d(a}\delta^{c}_{b)}\,. \label{rank10}
\ee 
Similarly we obtain the matrices in the one-form representation,
\be
(\Delta_{ab}^{(0,1)})_c{}^d = 2\eta_{ca}\delta^d_b\,, \quad \text{i.e.} \quad (r_{ab}^{(0,1)})_c{}^d = r_{ab}\delta^d_c+2\eta_{c[a}\delta^{d}_{b]} \quad \text{and} \quad
 (q_{ab}^{(1)})^c{}_d = q_{ab}\delta^d_c+2\eta_{c(a}\delta^{d}_{b)}\,. 
\ee
From these we can build the matrices for tensors of an arbitrary rank by simply taking the direct product of the above. For example, the
matrices for rank $(0,2)$ tensors are given as
\be
(\Delta_{ab}^{(0,2)})_c{}^d{}_e{}^f = (\Delta_{ab}^{(0,1)})_c{}^d \delta_e^f + \delta^d_c(\Delta_{ab}^{(0,1)})_e{}^f\,. \label{rank02}
\ee
Our convention is such that the gauge field $\bLambda$ is represented as $\bLambda = -\frac{1}{2}\bLambda^{ba}\Delta_{ab}$. 
Then, given for example a vector $V^a$, according to (\ref{rank10}) we have $\Diff V^a = \diff V^a + \bLambda^a{}_b V^b$.   
As another example, the constant $\eta_{ab}$ lives in the representation (\ref{rank02}), and thus we get
\be
\Diff\eta_{ab} = -\frac{1}{2}\bLambda^{ec} (\Delta_{ce}^{(0,2)})_a{}^d{}_b{}^f \eta_{df} = -2\bLambda_{(ab)}\,,
\ee
in agreement with (\ref{Qform}). 

Having reviewed the construction of tensor representations, we can finally proceed to spinors. The Lie derivative of a spinor field
$\psi$ along the vector $\bV$ is defined as on a metric manifold as
\be
\mathcal{L}_{\bV}\psi =  V^a\mathcal{D}_a\psi - \frac{1}{4}\mathcal{D}_{a}V_{b}\gamma^a\gamma^b\psi\,.
\ee
If $\bV$ is assumed to be a Killing vector of the metric whose covariant derivative $\mathcal{D}_a$ is, we have 
$\mathcal{D}_{a}V_{b} = \mathcal{D}_{[a}V_{b]}$. The Kosmann lift generalises the above formula for arbitrary vectors that need 
not be Killing vectors by imposing the antisymmetrisation. We have not imposed the antisymmetrisation, but it is easy to see that this 
the only difference this would make is that in the result (\ref{imag}) we would have $3\bQ+\bZ$ replaced by $\bZ$. The metric we consider
is the Minkowski metric of the tangent space, and thus the metric-covariant derivatives reduce to partial derivatives. 
The Lie derivatives of a spinor along the generating vectors of the GL become 
\be
\mathcal{L}_{\Sigma_{ab}}\psi = \Sigma_{ab}\psi - \frac{1}{2}\gamma_a\gamma_b\psi = \Sigma_{ab}\psi + \Delta_{ab}^{(\frac{1}{2})}\psi\,,
\ee
and thus, in accordance with (\ref{bGamma}), 
\be \label{spinorrep}
\Delta_{ab}^{(\frac{1}{2})} = -\frac{1}{2}\gamma_a\gamma_b\,, \quad \text{i.e.} \quad r_{ab}^{(\frac{1}{2})} = r_{ab} - \frac{1}{2}\gamma_{[a}\gamma_{b]} \quad \text{and} \quad
 q_{ab}^{(\frac{1}{2})} = q_{ab} + \frac{1}{4}\eta_{ab}\,. 
\ee
This completes our justification for the use of (\ref{spinorrep}) in the calculations.

\begin{center}
\begin{table}[h]
\begin{tabular}{|c| c| c| c| c| c |c |}
\hline
Transformation & Matrix  & Vector & $(\Delta^{(1,0)})^c{}_d$ & $(\Delta^{(0,1)})_c{}^d$ & $\Delta^{(\frac{1}{2})}$ & Potential \\
\hline
Translation & $\overset{+}{\gamma}{}_a  =  \frac{1}{2}\lp 1 + \gamma^5\rp \gamma_a$  & $\partial_a$ & $0$& $0$& $0$ & $\btau^a$ \\
Co-translation & $\overset{-}{\gamma}{}_a   =  \frac{1}{2}\lp 1 - \gamma^5\rp \gamma_a$ & $x^2\partial_a-2x_a x^b\partial_b$ &
$2\eta_{ad}x^c-4\delta^c_{[a}x_{d]}$ & $-2\eta_{ac}x^d+4\delta^d_{[a}x_{c]}$ & $x_a$ & $\bsigma^a$ \\
Rotation & $ -\frac{1}{2}\gamma_{[a}\gamma_{b]}$ & $r_{ab} = 2x_{[a}\partial_{b]}$ & $-2\eta_{d[a}\delta^c_{b]}$ &
$2\eta_{c[a}\delta^d_{b]}$ & $-\frac{1}{2}\gamma_{[a}\gamma_{b]}$ & $\bomega^{ab}$ \\
Dilation & $-\frac{1}{2}\gamma^5$ & $x^c\partial_c$ & $-\delta^c_d$ & $\delta^d_c$ & $2$ & $\bkappa$ \\
Identity & $1$ & $0$ & $0$ & $0$ & $0$ & $\bz$ \\
\hline
\end{tabular}
\caption{\label{table1} The elements of the centrally extended conformal group in terms of generating vectors and in terms of 16 4$\times$4 matrices, and the matrix representations corresponding to the former.}
\end{table}
\end{center}

The group SL(4,$\mathbbm{C}$) is the double cover of the group SO(6,$\mathbbm{C}$). The general linear algebra must thus be isomorphic to the
conformal algebra extended by including the central element. For curiosity, we shall check some properties of the representations in the 
conformal basis of the algebra. Some results are summarised in Table \ref{table1}. Comparing with our results, it looks like $\bQ$ and $\tilde{\bQ}$ correspond to the pieces $\bkappa$ and $x^a\bsigma_{a}$.
Finally, for whatever it might be good for, we could write down a spinor connection
\be
\Diff = \diff + \bt^a\overset{+}{\gamma}{}_a + \bsigma_a \overset{-}{\gamma}{}^a - \frac{1}{4}\bomega^{ab}\gamma_{[a}\gamma_{b]} - \frac{1}{2}\bkappa\gamma^5 + \bz\,,
\ee
and couple this connection into the Dirac action (\ref{dirac1}), to obtain its Hermitean version that survives in that action,
\ba
\Gamma^{\text{H}} & = & 
\frac{1}{4}\text{Re}\lp \tau^a{}_\mu+ \sigma^a{}_\mu\rp[\gamma^\mu,\gamma_a] + 
\frac{1}{2}\text{Re}\lp \tau^a{}_\mu - \sigma^a{}_\mu\rp\delta^\mu_a\gamma^5 
- \frac{i}{2}\epsilon^{abcd}\text{Re}(\omega_{a\mu b})\ie_c{}^\mu\gamma_d\gamma^5
+ \text{Re}(\kappa_\mu)\gamma^\mu\gamma^5 
\nonumber \\ 
& - &  \frac{1}{2}\text{Im}\lp \tau^a{}_\mu + \sigma^a{}_\mu\rp\delta^\mu_a
-  \frac{1}{4}\text{Im}\lp \tau^a{}_\mu - \sigma^a{}_\mu\rp\gamma^5 [\gamma^\mu,\gamma_a]
+ i\text{Im}(\omega^a{}_{\mu b})\ie_a{}^\mu\gamma^b
+ \text{Im}(z_\mu)\gamma^\mu\,.
\ea

\bibliography{spinQ}

\end{document}